# Supercell-enabled inverse design of boundary-state dispersion for group-velocity control in pseudomagnetic phononic crystals

Hua-Yang Chen[1], Nan-Xin Yu[1], Zhen-Hui Qin[1], Yan-Feng Chen[1,2,3], and Si-Yuan Yu[1,2,3]*

[1]National Laboratory of Solid-State Microstructures & Department of Materials Science and Engineering, Nanjing University, Nanjing 210093, China.

[2]Collaborative Innovation Center of Advanced Microstructures, Nanjing University, Nanjing 210093, China

[3]Jiangsu Key Laboratory of Artificial Functional Materials, Nanjing University, Nanjing, 210093, China

*corresponding author. e-mail: yusiyuan@nju.edu.cn

**Abstract:** Topological boundary states offer a platform for guided-wave propagation within bulk bandgaps. Their propagation characteristics depend not only on the spectral extent of the states but also on their dispersion within the operating bandwidth. A nearly linear dispersion provides more uniform group velocity across frequency, whereas a reduced slope enables slow-wave propagation. Despite advances in inverse design of prescribed dispersion relations and topological dispersion tuning, application-oriented characteristics of boundary-state dispersion have rarely been direct inverse-design objectives. Here, we develop a supercell-enabled, surrogate-assisted framework for the inverse design of boundary-state dispersion in a phononic crystal with pseudomagnetic fields. Area-preserving anisotropic deformation shifts the local Dirac points, and independent variation of 21 rows within the supercell provides distributed control over the boundary-state branch. A constraint-preserving surrogate reconstructs the dispersion from an initial frequency and positive frequency increments, with first- and second-difference losses introduced to capture local slope and curvature. Coupled with a genetic algorithm, the framework targets two prescribed-bandwidth design tasks: highly linear dispersion and slow-wave dispersion. For the former, finite-element verification yields relative group-velocity variations of 0.142% to 1.183% over prescribed bandwidths of 4% to 6%. For slow-wave designs, mean group velocities of 190 to 385 m/s are obtained over bandwidths of 0.2% to 0.8% while satisfying a 5% group-velocity variation limit. These results establish boundary-state dispersion as a direct inverse-design target beyond conventional bandwidth optimization, opening a flexible route to group-velocity control through application-oriented dispersion engineering over prescribed relative bandwidths.

## 1. Introduction

Phononic crystals tailor elastic-wave dispersion through structural periodicity, providing a versatile platform for bandgap engineering [1,2], wave manipulation [3–6], energy confinement and localization [7–10], and waveguiding [11–14]. Topological phononic crystals further exploit band degeneracies, symmetry breaking, and bulk–boundary correspondence to support topologically protected boundary states within bulk bandgaps [15,16]. In Dirac-type systems, structural modifications can lift Dirac degeneracies to open bandgaps or continuously shift Dirac points in momentum space. When the Dirac-point shift varies spatially, the low-energy effective description admits a valley-dependent pseudogauge potential and an associated pseudomagnetic field. This mechanism introduces additional structural degrees of freedom for continuously tailoring boundary-state dispersion and propagation characteristics [17–19]. However, the propagation characteristics of a boundary state depend not only on its existence and spectral extent, but also on the shape of its dispersion within the operating bandwidth. In particular, the slope and curvature of the dispersion determine the group velocity and its variation across frequency: a nearly linear dispersion provides more uniform group velocity across frequency [20], whereas a reduced dispersion slope enables slow-wave propagation and increased propagation delay [21,22]. These application-oriented requirements motivate direct control of boundary-state dispersion over prescribed frequency ranges, rather than focusing only on the existence or spectral extent of the boundary state.

Topology optimization [23–26], evolutionary algorithms [27–29], and machine learning [30–32] have been widely applied to the inverse design of phononic crystals and topological metamaterials [33,34], enabling structural optimization for prescribed bandgaps [35–37], target frequencies [38,39], and bandgaps for specific guided-wave modes [40,41]. Recent studies have further extended inverse design toward prescribed or customized dispersion relations in phononic crystals [35,42,43], while dispersion tuning and slow-wave behavior of boundary states have also been demonstrated in topological phononic systems [44]. Yet application-oriented characteristics of boundary-state dispersion, such as high linearity over a prescribed bandwidth or a reduced slope for slow-wave operation, have rarely been treated as direct inverse-design objectives. Unlike single-frequency or bandgap design, finite-bandwidth dispersion design requires coordinating the coupled effects of multiple structural degrees of freedom on an entire branch within a prescribed wavevector interval. Moreover, different parameter distributions can produce similar dispersion characteristics. The resulting non-unique mapping from target dispersion to structural parameters is difficult to resolve through direct single-valued regression. Group velocity introduces an additional challenge because it depends on the first derivative of the dispersion relation. Its variation over the operating band is therefore sensitive to local curvature and the resulting group-velocity extrema. If a surrogate model minimizes only pointwise frequency errors, small local deviations or violations of frequency ordering can be amplified during continuous reconstruction and differentiation. These considerations motivate combining forward surrogate prediction with optimization-based search to accommodate non-unique inverse solutions, while preserving frequency ordering and accurately representing dispersion slope and curvature.

In this paper, we develop a supercell-enabled, surrogate-assisted framework for the inverse design of boundary-state dispersion in a phononic crystal with pseudomagnetic fields. The system consists of an aluminum-alloy plate patterned with a triangular lattice of triangular through-holes. Area-preserving anisotropic deformation shifts the local Dirac points, while independently prescribed deformation parameters for 21 interior rows of unit cells within the supercell define a distributed design space for tailoring boundary-state dispersion. Using eigenfrequency data computed by the finite element method (FEM), we construct a constraint-preserving residual multilayer perceptron (ResMLP) that maps the 21 structural parameters to 25 boundary-state frequency samples. Reconstruction through positive frequency increments preserves the ordering of the predicted samples, while first- and second-difference losses improve the representation of dispersion slope and curvature. A five-model ensemble reduces variability associated with individual training runs. The ensemble is then coupled with a genetic algorithm to form a general surrogate-assisted inverse-design framework, which is demonstrated through two representative prescribed-bandwidth design tasks: highly linear dispersion and slow-wave dispersion. The dispersion relations and associated group-velocity characteristics of all selected candidate structures are independently recomputed using FEM for quantitative verification. By treating boundary-state dispersion itself as the direct design target, the framework extends inverse design beyond conventional discrete-frequency or bandwidth optimization toward group-velocity control through application-oriented dispersion engineering.

## 2. Phononic crystals with pseudomagnetic fields and boundary-state dispersion control

### 2.1. Unit-cell geometry and deformation-induced Dirac-point shifts

The system consists of an aluminum-alloy plate patterned with a triangular lattice of triangular through-holes, as shown in Fig. 1(a). The lattice constant is $a = 10\,mm$, the plate thickness is $h = 0.4a$, and the characteristic hole dimension is $r = 0.45a$. The density, Young's modulus, and Poisson's ratio are $\rho = 2700\,kg/m^3$, $E = 70\,GPa$, and $\nu = 0.33$, respectively. The transverse and propagation directions are defined as $x$ and $y$, respectively. Taking the undeformed triangular hole as the reference configuration, reciprocal scaling factors are applied along these two directions:

$$l_x(s) = \frac{l_{x0}}{1+s}, \qquad l_y(s) = (1+s)l_{y0}. \tag{1}$$

Here, $s \in [-0.1, 0.1]$ is the dimensionless deformation parameter, and $l_{x0}$ and $l_{y0}$ are the corresponding reference dimensions. This transformation changes the anisotropy of the hole while preserving its area.

Fig. 1(b) shows the transverse arrangement of the supercell and the local band structures of representative unit cells. The supercell comprises 21 interior rows of deformable unit cells bounded by gapped unit cells with opposite hole rotations. For the unrotated interior cells, the three central band-structure panels correspond to $s = 0.1$, 0, and -0.1, from left to right. Relative to the undeformed configuration, positive and negative deformations shift the Dirac point toward smaller and larger $k_y$, respectively, with only a small change in the Dirac frequency. Thus, the area-preserving anisotropic deformation primarily controls the momentum-space position of the Dirac point.

The leftmost and rightmost band-structure panels correspond to unit cells with hole rotations of $-30°$ and $+30°$, respectively. Both rotations lift the Dirac degeneracy and open bandgaps with similar frequency ranges and widths. In an effective Dirac description, the oppositely rotated configurations are represented by mass terms of opposite signs, providing the conditions for boundary-state formation. The two gapped regions inhibit outward transverse propagation within the target frequency range, confining the boundary modes primarily to the interior design region and its vicinity. With these outer regions fixed, independently adjusting the interior deformation parameters enables control of both boundary-state dispersion and the transverse energy distribution.

These local band-structure changes can be described qualitatively using an effective model near the Dirac points. In the selected modal basis, the effective Hamiltonian is written as

$$H_\tau = f_D(s)I + \frac{v_y}{2\pi}[q_y - A_{\tau,y}(s)]\sigma_y + \tau\frac{v_x}{2\pi}q_x\sigma_x + M(\theta)\sigma_z. \tag{2}$$

Here, $\tau = \pm 1$ labels the two valleys related by time-reversal symmetry, $f_D(s)$ is the deformation-dependent Dirac frequency, and $\boldsymbol{q}$ is the wavevector measured relative to the undeformed reference Dirac point in each valley. The parameters $v_x$ and $v_y$ are the characteristic Dirac-cone velocities along the two directions, $\sigma_i$ are the Pauli matrices acting in the two-mode subspace, and $I$ is the identity matrix. The mass term $M(\theta)$ arises from rotation of the triangular hole. The unrotated configuration satisfies $M(0) = 0$, whereas $\theta = \pm 30°$ produces nonzero mass terms with opposite signs.

Let $k_D(s)$ denote the Dirac-point position in the valley shown in Fig. 1(b), corresponding to $\tau = +1$. Its displacement along $k_y$ relative to the reference configuration is

$$\delta k_D(s) = k_D(s) - k_D(0). \tag{3}$$

The normalized Dirac-point position in Fig. 1(b) is $\kappa_D(s) = k_D(s)/[4\pi/(3a)]$. In Eq. (2), $q_y$, $\delta k_D$, and the gauge potential have the same wavevector units. Accounting for the time-reversal relation between the two valleys, the deformation-induced Dirac-point shift can be represented by a valley-dependent gauge potential:

$$A_{\tau,y}(s) = \tau \delta k_D(s). \tag{4}$$

When the deformation profile can be approximated by a slowly varying continuous function $s(x)$, the gauge potential takes the form $\boldsymbol{A}_\tau(x) = [0, A_{\tau,y}(x)]$. Its curl defines a pseudomagnetic field normal to the plate:

$$B_{\tau,z}(x) = [\nabla \times \boldsymbol{A}_\tau(x)]_z = \frac{\partial A_{\tau,y}(x)}{\partial x}. \tag{5}$$

Because both the gauge potential and the resulting pseudomagnetic field have opposite signs in the two valleys, the system preserves time-reversal symmetry. Eqs. (2)–(5) provide a qualitative connection between structural deformation, Dirac-point shifts, and pseudomagnetic fields. The full boundary-state dispersion is obtained from FEM eigenfrequency analysis of the supercell.

### 2.2. Distributed deformation parameters and tunable boundary states

To introduce distributed structural degrees of freedom, 21 rows of unrotated, deformable unit cells are arranged along the transverse $x$-direction. Each row is assigned an independent deformation parameter $s_i$, giving the design vector

$$\boldsymbol{s} = [s_1, s_2, \cdots, s_{21}]^{\mathrm{T}}, \quad -0.1 \leq s_i \leq 0.1. \tag{6}$$

During inverse design, the two outer regions of rotated unit cells and their associated mass terms remain fixed; only the deformation parameters of the 21 interior rows are varied. By shifting the local Dirac points, these parameters modify the transverse distribution of $A_{\tau,y}$ and thereby influence boundary-state dispersion. For deformation profiles that vary slowly along $x$, the corresponding pseudomagnetic field can be interpreted using Eq. (5). A uniform deformation, in contrast, produces no gauge-potential gradient within the design region. General combinations of the 21 discrete parameters are treated as distributed structural design variables and are not required to represent a continuous or uniform pseudomagnetic field.

Boundary modes at different wavevectors exhibit different transverse spatial distributions, and their eigenfrequencies are jointly influenced by multiple rows of unit cells. The effects of the deformation parameters on dispersion are therefore spatially coupled. Coordinated adjustment of $s_i$ can modify not only the frequency position of the branch but also its local slope and curvature.

Figs. 1(c) and 1(d) compare two representative parameter profiles. The uniform configuration with $s_i = 0$ serves as a reference. When $s_i$ gradually changes from positive to negative values across the transverse direction, the boundary-state dispersion exhibits appreciable changes in slope and curvature relative to this reference. Because the outer gapped regions remain unchanged, this comparison demonstrates that the interior deformation profile provides effective control over boundary-state dispersion.

In Figs. 1(c) and 1(d), the supercell eigenmodes at 76.5 kHz are shown above the dispersion curves, while the fields excited by a plane wave at the same frequency are shown below. The field colors represent normalized energy density. White dashed lines mark the interfaces between the design region and the two outer topological insulators. The eigenmodes are primarily localized between these interfaces and decay into the outer regions. The driven fields likewise show energy propagation along the design region, confirming that the plotted dispersion branches correspond to boundary states that can be excited and support transmission at this frequency.

Compared with the broader transverse distribution in the uniform configuration, the gradient configuration concentrates energy more strongly within the design region. The eigenmode and transmission fields exhibit consistent trends. These results indicate that the deformation parameters control both boundary-state dispersion and the transverse energy distribution within the propagation channel. The surrogate model developed below learns the mapping from the 21 deformation parameters to boundary-state dispersion using full FEM results and subsequently supports inverse design.

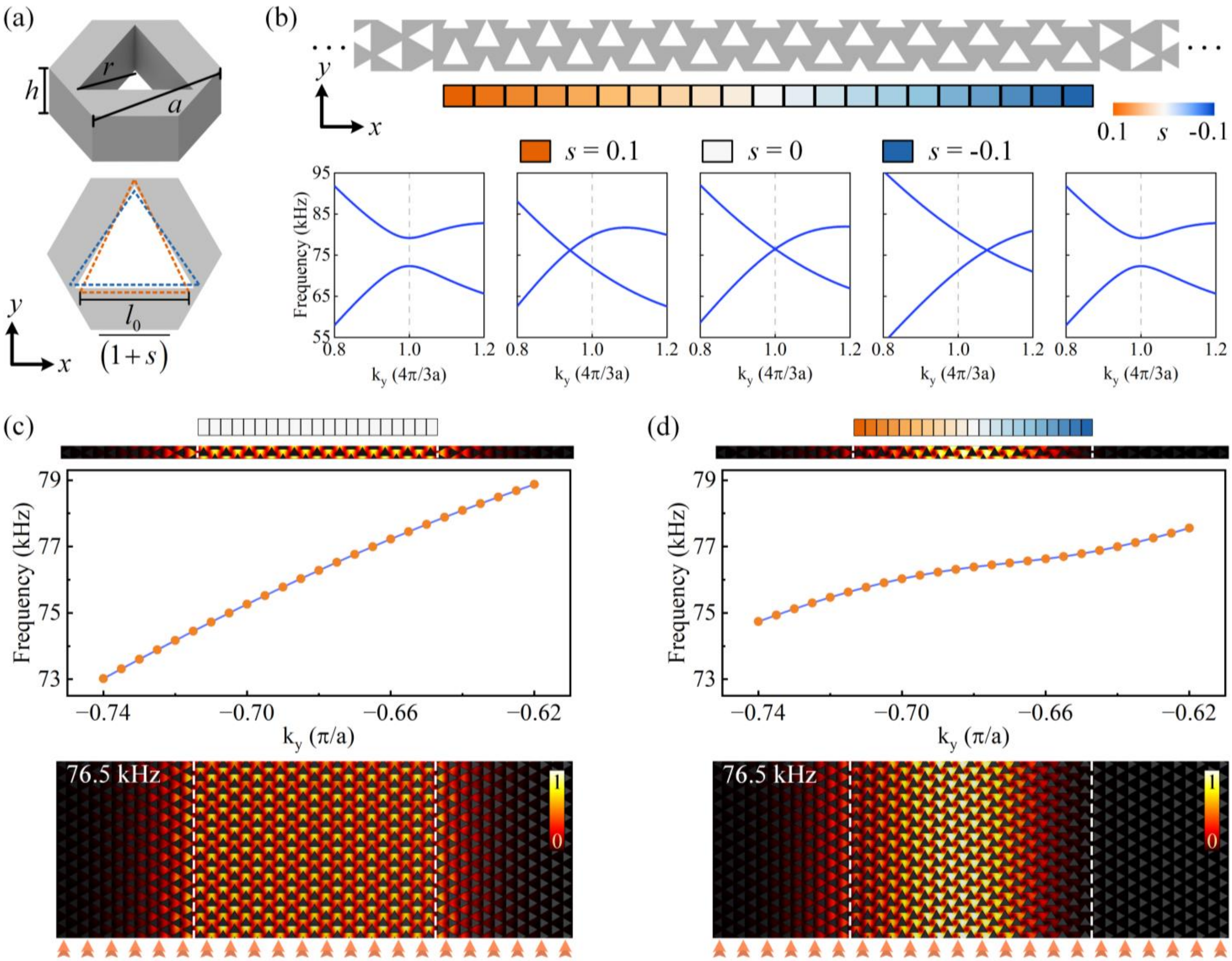

**Fig. 1.** Phononic crystal with pseudomagnetic fields and boundary-state dispersion control. (a) Unit-cell geometry and area-preserving anisotropic deformation. (b) Supercell arrangement and local band structures. The three central panels correspond to $s = 0.1$, 0, and -0.1; the outer panels correspond to gapped cells with hole rotations of $-30°$ and $+30°$. (c) Uniform profile with $s_i = 0$. (d) Transverse gradient profile from positive to negative $s_i$. In (c) and (d), the parameter profile, supercell eigenmode, boundary-state dispersion, and plane-wave-excited transmission field are shown from top to bottom. All field maps show normalized energy density at 76.5 kHz. White dashed lines mark the design-region interfaces, and orange arrows indicate the incident-wave direction.

### 2.3. Finite element calculations and dataset

Boundary-state dispersion is obtained through FEM eigenfrequency analysis with Bloch-periodic boundary conditions. The normalized wavevector is defined as $m = k_y a/\pi$. The target branch is sampled over $m \in [-0.74, -0.62]$ with an increment of 0.005, yielding 25 wavevector points.

FEM eigenfrequency calculations are performed for structural configurations within the prescribed parameter bounds, and the target boundary-state branches are identified and selected from the resulting band structures. The 25 frequency samples over $m \in [-0.74, -0.62]$ are then paired with the corresponding 21-dimensional structural parameter vector. A total of 120000 samples are obtained and divided into training, validation, and independent test sets at a ratio of 8:1:1. This dataset is used to establish the forward mapping from the 21 structural parameters to the 25-point boundary-state dispersion.

## 3. Constraint-preserving residual surrogate model

### 3.1. Input representation and residual feature extraction

The surrogate model takes the structural parameter vector $\boldsymbol{s} \in \mathbb{R}^{21}$ as input and predicts 25 discrete frequencies along the corresponding boundary-state dispersion branch. Each input parameter is standardized using a z-score transformation to account for differences in statistical scale:

$$\tilde{s}_i = \frac{s_i - \mu_i}{\sigma_i}, \qquad i = 1, \cdots, 21. \tag{7}$$

Here, $\mu_i$ and $\sigma_i$ are the mean and standard deviation of the $i$th structural parameter, calculated exclusively from the training set. These statistics are subsequently fixed and applied to the validation set, test set, and inverse-design calculations, preventing information leakage during preprocessing.

The standardized 21-dimensional input is first mapped to a 256-dimensional feature space through a fully connected layer, followed by layer normalization (LayerNorm) and a SiLU activation. The network backbone consists of six pre-normalized residual blocks, each with a width of 256. The $l$th residual block is expressed as

$$\boldsymbol{h}_{l+1} = \boldsymbol{h}_l + \boldsymbol{W}_{l,2}\phi\left[\boldsymbol{W}_{l,1}\mathrm{LN}(\boldsymbol{h}_l) + \boldsymbol{b}_{l,1}\right] + \boldsymbol{b}_{l,2}. \tag{8}$$

Here, $\phi(\cdot)$ denotes the SiLU activation, and $\mathrm{LN}(\cdot)$ denotes layer normalization. The residual connection provides a direct path for feature propagation, allowing each block to learn a nonlinear correction to the existing representation. Normalizing the features before the residual transformation helps stabilize gradient propagation through the network.

The residual multilayer perceptron represents the nonlinear mapping between the 21 deformation parameters and the discrete boundary-state frequencies. However, accurate pointwise frequency predictions alone do not ensure reliable derivative estimates for subsequent group-velocity evaluation. A constraint-preserving output parameterization and derivative-aware losses are therefore introduced to maintain the required frequency ordering and improve the representation of local dispersion slope and curvature.

### 3.2. Discrete monotonic outputs and derivative-aware losses

Conventional multi-output regression networks predict the 25 frequency samples without explicitly enforcing their ordering, potentially introducing local non-monotonicity into the predicted branch. Such ordering errors can affect continuous dispersion reconstruction and the subsequent differentiation used to calculate group velocity. To preserve the increasing frequency sequence of the target branch, the output is parameterized by an initial frequency and positive frequency increments. The complete branch is reconstructed through cumulative summation:

$$\hat{f}_1 = f_{\text{off}} + f_{\text{scale}} z_0, \tag{9}$$

$$\Delta\hat{f}_i = f_{\text{scale}} \text{softplus}(z_i) + \varepsilon_f, \qquad i = 1, \cdots, 24, \tag{10}$$

$$\hat{f}_{i+1} = \hat{f}_1 + \sum_{j=1}^{i} \Delta\hat{f}_j\,. \tag{11}$$

Here, $z_0$ and $z_i$ are the network outputs associated with the initial frequency and frequency increments, respectively. The offset $f_{\text{off}}$ and scale factor $f_{\text{scale}}$ are used to restore dimensional frequencies. They correspond to the mean initial frequency and the overall frequency standard deviation of the training set, respectively. The positive minimum increment $\varepsilon_f$ prevents adjacent frequency samples from coinciding. Since $\text{softplus}(z_i) > 0$, Eqs. (9)–(11) guarantee $\hat{f}_{i+1} > \hat{f}_i$ at the 25 sampled wavevectors. The initial frequency controls the overall frequency position of the branch, whereas the positive increments describe its variation between adjacent wavevector samples.

Frequency ordering alone does not ensure that the predicted branch reproduces the local variations of the target dispersion. Because group velocity depends on the first derivative and local curvature is characterized by the second derivative, first- and second-difference terms are incorporated into the loss function. For a discrete frequency sequence, these differences are defined as

$$\Delta f_i = f_{i+1} - f_i, \tag{12}$$

$$\Delta^2 f_i = f_{i+2} - 2f_{i+1} + f_i. \tag{13}$$

Writing the frequency, first difference, and second difference collectively as $q \in \{f, \Delta f, \Delta^2 f\}$, the corresponding loss term is

$$L_q = \frac{1}{N_q} \sum_{n=1}^{N_q} \left( \frac{\hat{q}_n - q_n}{\sigma_q} \right)^2. \tag{14}$$

Here, $N_q$ is the number of entries included in the corresponding loss, and $\sigma_q$ is the training-set standard deviation of the relevant quantity. This normalization balances the numerical scales of the loss terms. All standard deviations are calculated exclusively from the training set and remain fixed thereafter. The total training loss is

$$L = L_f + 0.5L_{\Delta f} + 0.2L_{\Delta^2 f}. \tag{15}$$

The frequency loss $L_f$ penalizes pointwise prediction errors, while $L_{\Delta f}$ and $L_{\Delta^2 f}$ penalize errors in adjacent frequency increments and their variations, respectively. Thus, Eq. (15) accounts for first- and second-order variations of the discrete dispersion branch in addition to pointwise frequency accuracy, providing more consistent inputs for continuous reconstruction and group-velocity evaluation.

The models are trained using the AdamW optimizer with an initial learning rate of $10^{-3}$, a weight decay of $10^{-5}$, and a batch size of 2048. Gradient norms are clipped at 5. Training runs for a maximum of 2000 epochs and stops early if the validation loss fails to improve by at least $10^{-6}$ for 100 consecutive epochs. Five models with identical architectures are trained independently using different random seeds and subsequently combined into an ensemble.

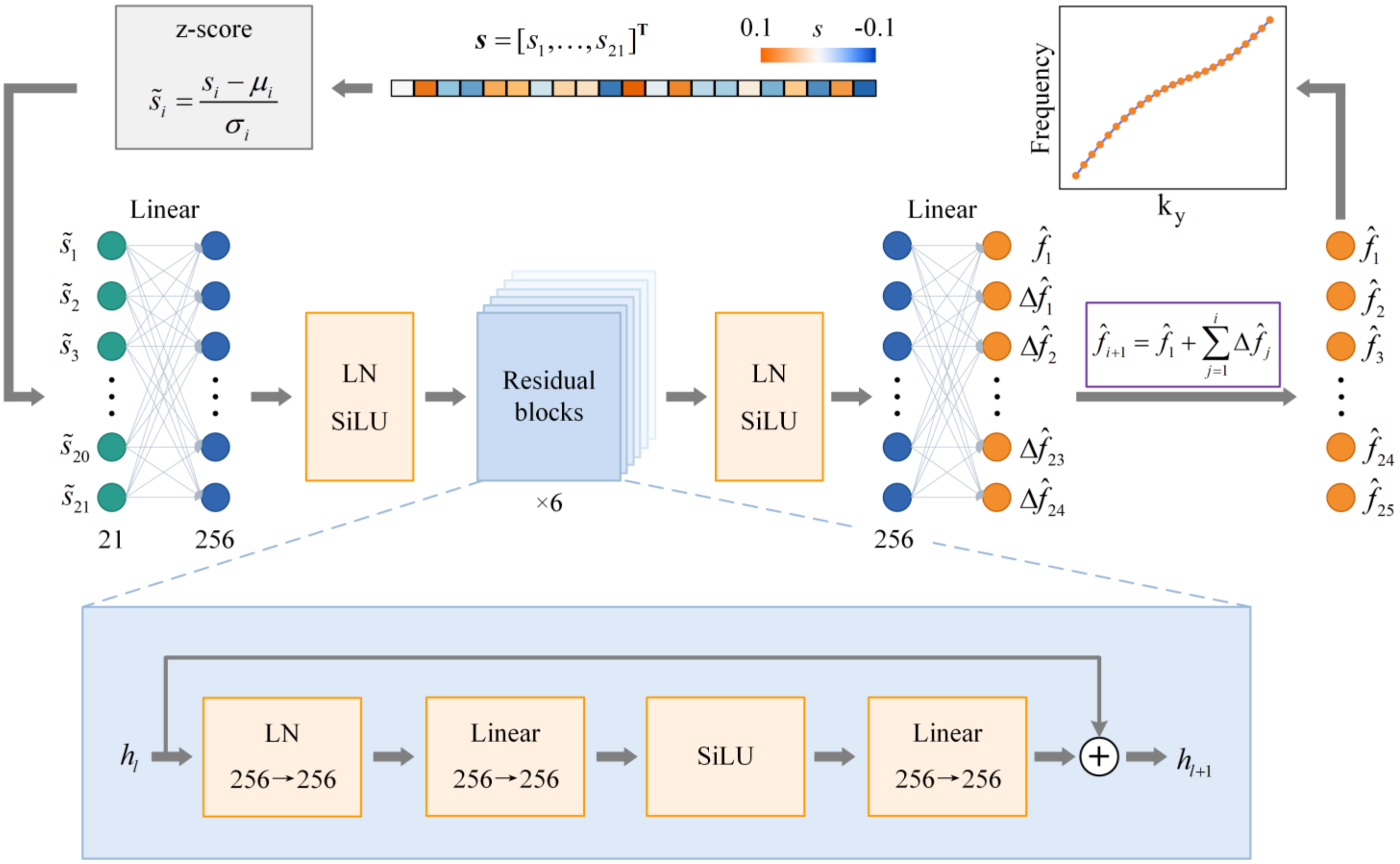

**Fig. 2.** Architecture of the constraint-preserving residual surrogate model. The 21 deformation parameters are standardized using training-set statistics and mapped to a 256-dimensional feature space. The backbone comprises six pre-normalized residual blocks of width 256. The output heads predict an initial frequency and 24 positive increments, which are cumulatively summed to reconstruct 25 strictly increasing boundary-state frequencies. The lower panel details the residual block, including LayerNorm, fully connected layers, SiLU activation, and the residual connection.

### 3.3. Five-model ensemble and forward-prediction assessment

To reduce variability associated with individual training runs, predictions from the five independently trained models are averaged. Each member predicts 25 discrete frequencies, and the ensemble prediction at the $i$th wavevector is defined as

$$\bar{f}_i = \frac{1}{5} \sum_{r=1}^{5} \hat{f}_i^{(r)}. \tag{16}$$

Here, $\hat{f}_i^{(r)}$ is the frequency predicted by the $r$th member at the $i$th wavevector, and $\bar{f}_i$ is the ensemble mean. The ensemble mean is used for both forward-performance assessment and inverse design. Ensembling is employed to reduce training-induced variability, rather than to identify out-of-distribution samples or quantify predictive uncertainty.

Fig. 3(a) shows the training and validation losses of the five ensemble members. The losses decrease progressively and stabilize without a pronounced, persistent divergence between training and validation curves, indicating stable convergence. Forward-prediction performance is further evaluated on the independent test set.

The ensemble mean defined in Eq. (16) is assessed using 12000 test samples. Fig. 3(b) compares the predicted frequencies with the FEM results. The points lie close to the identity line, yielding a coefficient of determination of $R^2 = 0.999985$. The frequency mean absolute error (MAE) and root mean square error (RMSE) are 4.593 Hz and 6.275 Hz, respectively, demonstrating accurate prediction of boundary-state dispersion from the structural parameters.

Fig. 3(c) presents the pointwise frequency-error distribution, which is concentrated near zero and has a 95% empirical percentile interval of [-11.6, 12.6] Hz. To assess derivative-level accuracy, group velocities are calculated using the same seventh-order Chebyshev fitting and differentiation procedure employed in inverse design. The resulting group-velocity errors have a 95% empirical percentile interval of [-3.41, 3.75] m/s, as shown in Fig. 3(d). The first-difference MAE and RMSE are 0.301 Hz and 0.616 Hz, respectively, while the corresponding second-difference errors are 0.072 Hz and 0.431 Hz. These results indicate that the surrogate preserves the local variations of the discrete dispersion branch. Applying the same reconstruction and differentiation procedure to the predicted and FEM dispersions also yields small group-velocity discrepancies, supporting the use of the surrogate for rapid dispersion prediction within the investigated design domain.

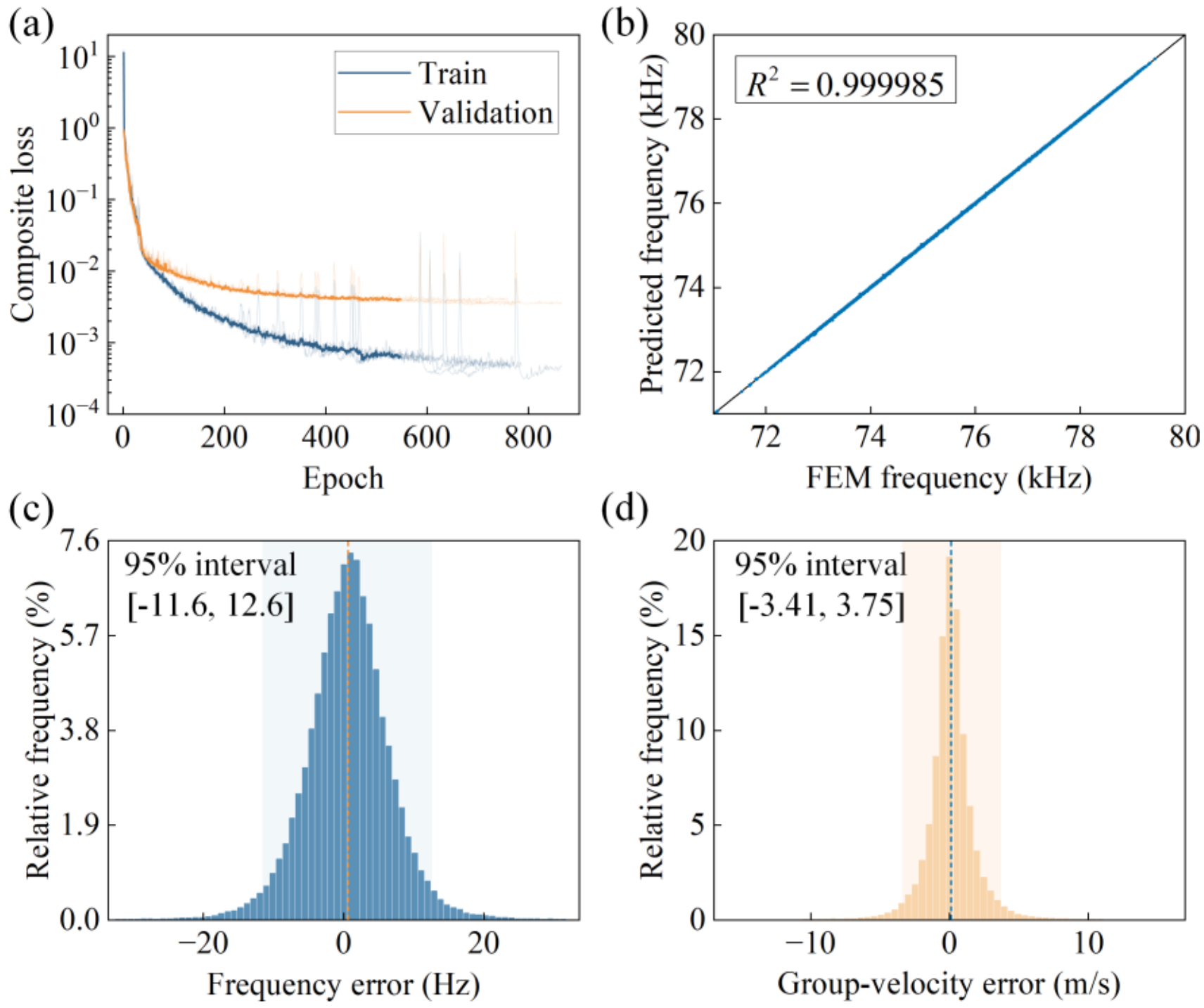

**Fig. 3.** Training and independent test results of the constraint-preserving residual surrogate model. (a) Training and validation losses of the five ensemble members; thin lines represent individual models, and thick lines indicate the overall trends. (b) Pointwise comparison between ensemble-predicted and FEM frequencies on the independent test set; the black diagonal denotes exact agreement. (c) Frequency-error distribution and its 95% empirical percentile interval. (d) Group-velocity-error distribution obtained by applying identical seventh-order Chebyshev fitting and differentiation to the predicted and FEM dispersions. Dashed lines indicate the centers of the error distributions, and shaded regions denote the corresponding 95% empirical percentile intervals.

# 4. Surrogate-assisted inverse design

## 4.1. Continuous dispersion reconstruction and group-velocity evaluation

For a given structural parameter vector, the five surrogate models predict 25 discrete frequencies, which are averaged according to Eq. (16). Because group velocity depends on the first derivative of the dispersion relation, these discrete predictions must be converted into a continuously differentiable representation. The normalized wavevector interval $m \in [-0.74, -0.62]$ is therefore mapped linearly to $\xi \in [-1, 1]$, and a seventh-order Chebyshev polynomial is fitted by least squares:

$$f_{\text{fit}}(\xi) = \sum_{n=0}^{7} c_n T_n(\xi) . \tag{17}$$

Here, $T_n(\xi)$ is the $n$th-order Chebyshev polynomial, and $c_n$ is the corresponding fitting coefficient.

Using $m = k_y a/\pi$ and $\omega = 2\pi f$, the group velocity is expressed as

$$v_g(m) = \frac{\partial \omega}{\partial k_y} = 2a \frac{df_{\text{fit}}}{dm}. \tag{18}$$

The fitted curve is evaluated at 801 equally spaced values of $m$ to assess group-velocity characteristics over different frequency windows. For a contiguous window bounded by frequencies $f_L$ and $f_R$, the relative bandwidth is defined as

$$\beta = \frac{2(f_R - f_L)}{f_R + f_L}. \tag{19}$$

All contiguous windows satisfying the prescribed bandwidth are examined. Within each window, the mean group velocity over the corresponding wavevector interval and the relative group-velocity variation are calculated as

$$\bar{v}_g = 2a \frac{f_R - f_L}{m_R - m_L}, \tag{20}$$

$$R_v = \frac{v_{g,\max} - v_{g,\min}}{\bar{v}_g}. \tag{21}$$

Here, $m_L$ and $m_R$ are the wavevector coordinates corresponding to the window endpoints, and $v_{g,max}$ and $v_{g,min}$ are the maximum and minimum group velocities within the window. The discrete monotonicity constraint preserves the ordering of the predicted frequency samples, providing a consistent basis for subsequent reconstruction and differentiation. A smaller $R_v$ indicates more uniform group velocity within the selected frequency window.

### 4.2. Design objectives and genetic search

Although structural deformation provides control over boundary-state dispersion, no direct analytical mapping is available between the structural parameters and the target group-velocity characteristics. The design variables jointly influence the full dispersion branch, while the objectives depend on group velocity, its extrema, and its mean over a continuous frequency window. Evaluating these quantities directly through FEM during genetic search would require extensive repeated eigenfrequency calculations. The constraint-preserving surrogate is therefore used for population evaluation, followed by independent FEM verification of the final candidate structures.

To demonstrate the flexibility of the framework, we consider two representative prescribed-bandwidth design objectives. The first targets highly linear dispersion, quantified by uniform group velocity over a finite bandwidth. Differences in group velocity between frequency components produce differential group delay and can cause wave-packet broadening. Minimizing $R_v$ therefore promotes more uniform propagation within the prescribed bandwidth. Since a strictly increasing dispersion branch can still exhibit substantial group-velocity variation, the optimization problem is formulated as

$$\min_{s,m_L,m_R} R_v, \quad s.t. \quad -0.1 \le s_i \le 0.1, \quad \beta(m_L, m_R) = \beta_0, \quad v_g > 0. \tag{22}$$

The target relative bandwidth $\beta_0$ is set to 4%, 5%, or 6%. For each candidate structure, the fitness is determined by the smallest $R_v$ among all windows satisfying the bandwidth requirement, without imposing an additional constraint on mean group velocity. This objective evaluates the ability of structural deformation to suppress group-velocity variation over a prescribed bandwidth.

The second representative objective targets slow-wave dispersion by minimizing the mean group velocity. Reducing group velocity increases group delay per unit propagation distance and extends wave–structure interaction time within an ideal lossless model. However, a locally low group velocity may be accompanied by substantial variation across the operating window. The second problem therefore minimizes mean group velocity while constraining its relative variation:

$$\min_{s,m_L,m_R} \bar{v}_g, \quad s.t. \quad -0.1 \le s_i \le 0.1, \quad \beta(m_L, m_R) = \beta_0, \quad v_g > 0, \quad R_v \le 0.05. \tag{23}$$

The target relative bandwidth $\beta_0$ is set to 0.2%, 0.4%, 0.6%, 0.8%, or 1.0%. Windows that violate $R_v \le 0.05$ are penalized and ranked below feasible windows. This objective thus seeks a finite-bandwidth slow-wave window with controlled group-velocity variation, rather than an extremely low group velocity at a single frequency.

Both problems are solved using a genetic algorithm. Each generation contains 128 candidate structures, with eight elite individuals retained. Tournament selection, crossover, and mutation are used to generate the next population. The design variables are normalized before optimization, and the mutation standard deviation is set to 0.05 in the normalized space, corresponding to approximately 0.01 in the deformation parameters. The search runs for a maximum of 5000 generations and terminates early if the best objective value shows no effective improvement for 300 consecutive generations. For the first objective, the candidate with the smallest $R_v$ is retained. For the second, the feasible candidate with the lowest mean group velocity is retained.

The term "optimal design" refers here to the best candidate obtained under the current surrogate model, parameter bounds, and genetic-algorithm settings, rather than a guaranteed global optimum. All final candidates are independently recomputed using FEM to obtain their 25-point boundary-state dispersions. The resulting curves are assessed using the same Chebyshev fitting and group-velocity evaluation procedures applied to the surrogate predictions.

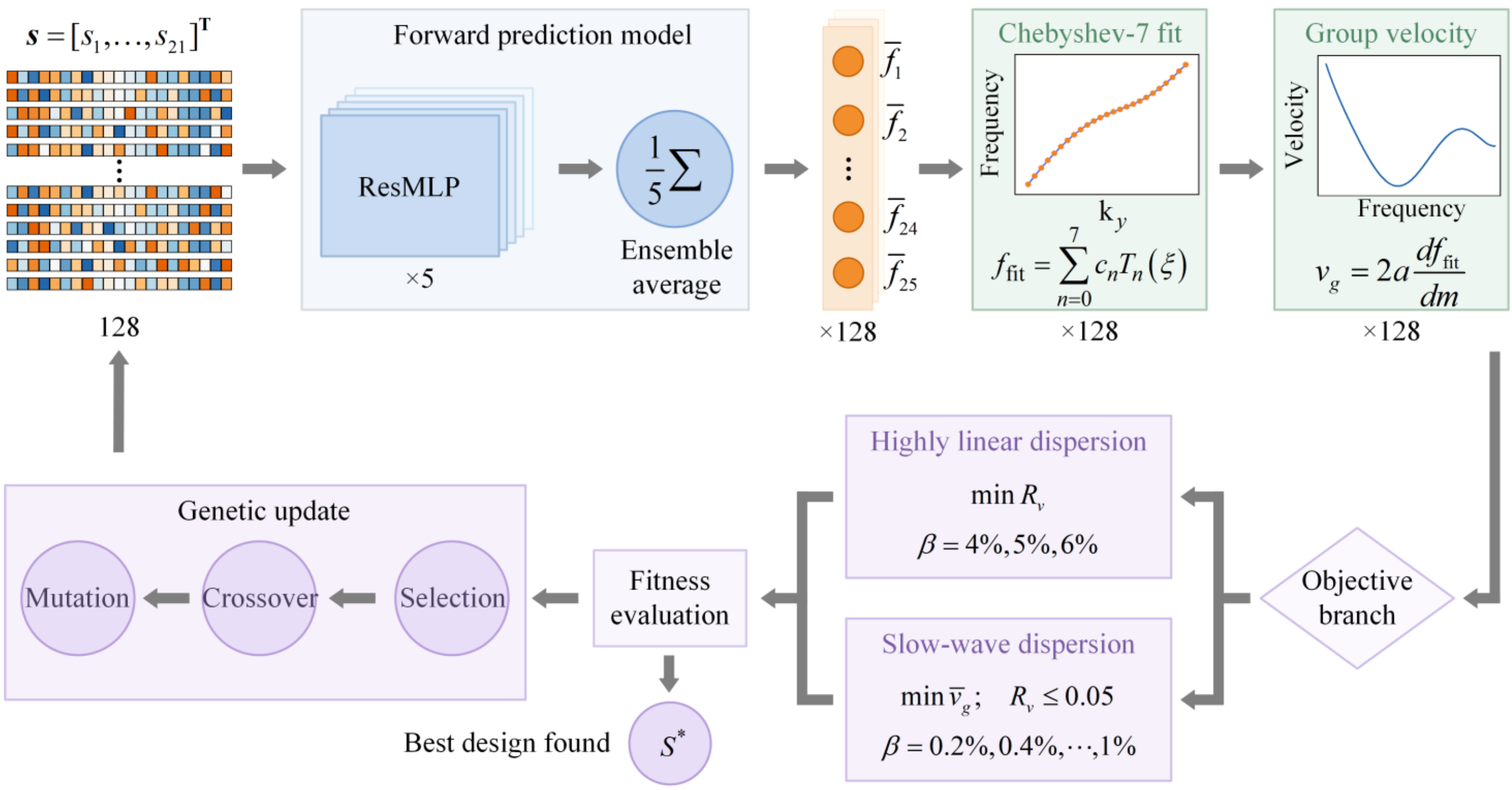

**Fig. 4.** Surrogate-assisted inverse design of boundary-state dispersion. Each generation contains 128 candidate structures, each defined by 21 deformation parameters. Five-model ensemble prediction, seventh-order Chebyshev fitting, and group-velocity calculation are followed by fitness evaluation for either finite-bandwidth group-velocity uniformity or low mean group velocity subject to a uniformity constraint. Selection, crossover, and mutation generate subsequent populations, while the best candidate is retained throughout the search. Final dispersion relations and group-velocity metrics are independently verified using FEM.

## 5. Results and discussion

### 5.1. Group-velocity uniformity over finite bandwidths

Fig. 5 presents the inverse-design results for target relative bandwidths of $\beta = 4\%$, 5%, and 6%. The corresponding 21-dimensional deformation profiles are shown in Figs. 5(a), 5(d), and 5(g). All three configurations exhibit nonuniform deformation, with substantial differences between the parameter profiles obtained for different bandwidths. These designs tailor the boundary-state dispersion through coordinated adjustment of the deformation parameters across multiple rows. Figs. 5(b), 5(e), and 5(h) compare the surrogate-predicted dispersion relations with those independently recomputed using FEM, showing good overall agreement. The corresponding group velocities are compared in Figs. 5(c), 5(f), and 5(i), where the gray regions indicate the target bandwidth windows. FEM verification yields mean group velocities of 882, 836, and 791 m/s for $\beta = 4\%$, 5%, and 6%, respectively, with corresponding relative group-velocity variations of $R_v = 0.1423\%$, 0.5769%, and 1.1826%.

As the target bandwidth increases from 4% to 6%, the evaluation window broadens and $R_v$ increases from 0.1423% to 1.1826%. This trend illustrates the trade-off between operating bandwidth and group-velocity uniformity: maintaining a nearly constant group velocity becomes more demanding over a wider frequency range. The frequency RMSEs between the surrogate predictions and FEM results are 3.348, 4.744, and 21.646 Hz for the three designs, respectively. Although the prediction error increases for the 6% bandwidth design, the surrogate still reproduces the overall dispersion trend. The FEM-recomputed value of $R_v = 1.1826\%$ confirms that this design maintains relatively uniform group velocity throughout the target window.

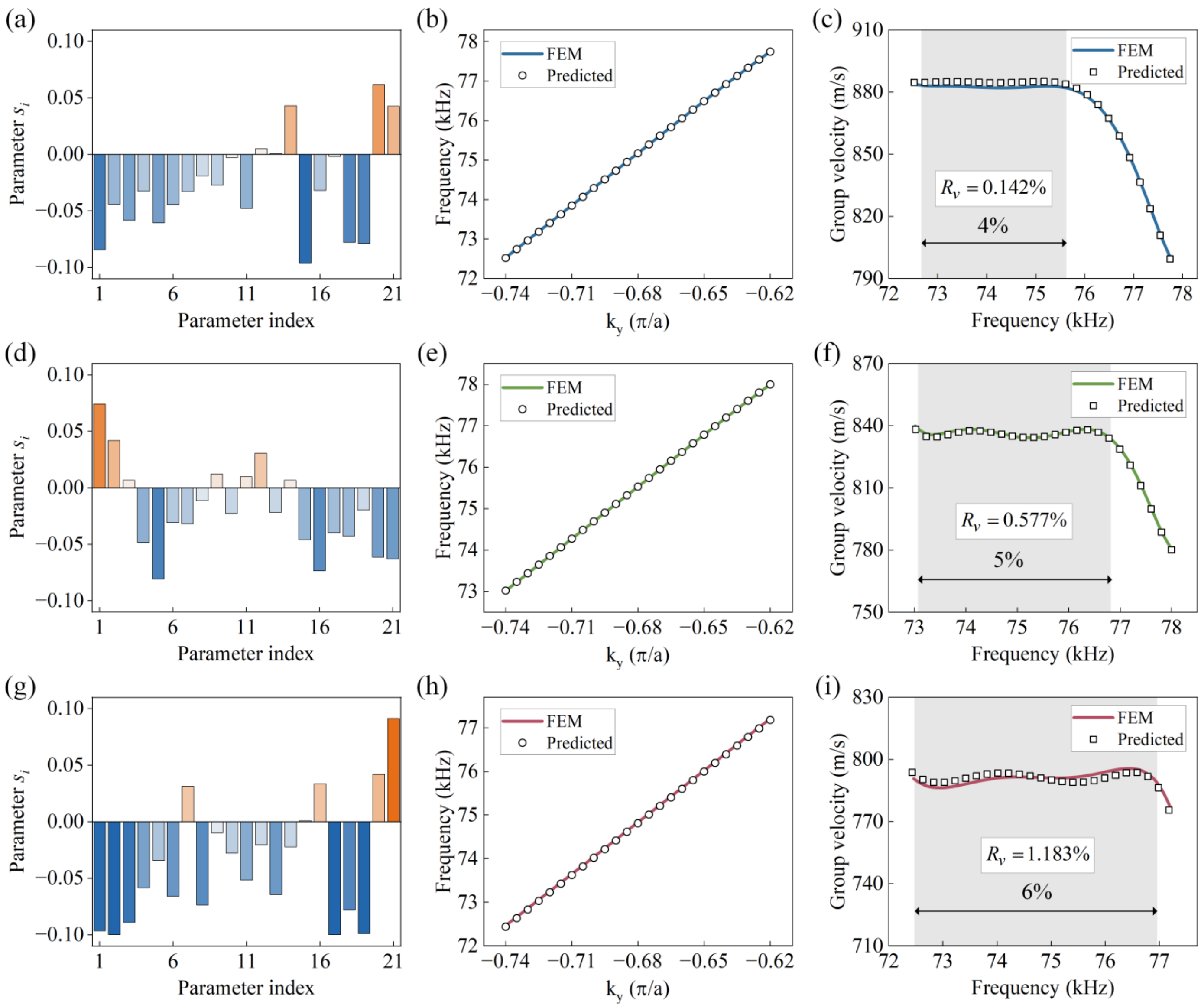

**Fig. 5.** Inverse design of group-velocity uniformity at different target relative bandwidths. Each row shows the 21-dimensional deformation profile, surrogate-predicted and FEM-recomputed boundary-state dispersions, and corresponding group velocities. Solid lines denote FEM results, markers denote surrogate predictions, and gray regions indicate the target bandwidth windows. (a–c) $\beta = 4\%$, with FEM-recomputed $R_v = 0.142\%$. (d–f) $\beta = 5\%$, with $R_v = 0.577\%$. (g–i) $\beta = 6\%$, with $R_v = 1.183\%$.

### 5.2. Low mean group velocity subject to a uniformity constraint

Fig. 6 presents the low-mean-group-velocity designs for target relative bandwidths of $\beta = 0.2\%$, 0.4%, 0.6%, 0.8%, and 1.0%. Figs. 6(a), 6(d), 6(g), 6(j), and 6(m) show the corresponding deformation profiles and plane-wave-excited energy-density fields at representative frequencies. The surrogate-predicted and FEM-recomputed dispersion relations are compared in Figs. 6(b), 6(e), 6(h), 6(k), and 6(n). The associated group velocities and target bandwidth windows are shown in Figs. 6(c), 6(f), 6(i), 6(l), and 6(o). Compared with the uniform-group-velocity designs in Section 5.1, these configurations contain more deformation parameters at or near the bounds $s_i = \pm 0.1$, together with contiguous groups of parameters sharing the same sign. Such profiles reflect the coordinated adjustment of deformation across different positions to reduce the dispersion slope.

The surrogate and FEM dispersion curves are generally close, and the principal group-velocity trends are reproduced. Nevertheless, discrepancies remain in the low-group-velocity regions and adjacent frequency ranges for some configurations. FEM verification gives mean group velocities of 190, 256, 326, 385, and 408 m/s for the five target bandwidths, respectively. The corresponding values of $R_v$ are 3.560%, 4.978%, 4.691%, 4.404%, and 6.102%. The first four designs satisfy the prescribed constraint $R_v \leq 5\%$, demonstrating that the surrogate can identify candidate structures combining low mean group velocity with controlled variation. The FEM result for the 1.0% bandwidth design, however, exceeds the prescribed limit.

To examine the driven propagation characteristics of the optimized structures, Fig. 6 includes plane-wave-excited fields at representative frequencies below each deformation profile. The field colors represent normalized energy density, and the white dashed lines mark the interfaces between the design region and the two outer topological insulators. In all five configurations, energy is primarily concentrated between these interfaces, forming a transversely confined propagation channel along the design region. These fields provide complementary evidence that the boundary states can be excited and support propagation at the selected frequencies. The transverse energy distributions differ between configurations at their respective representative frequencies. In Fig. 6(d), energy is concentrated mainly on the left side of the design region, whereas energy in the other configurations is concentrated mainly in the central-to-right portion of the design region. Thus, the propagation modes associated with the low-group-velocity designs do not exhibit a fixed transverse distribution; their spatial profiles depend jointly on the structural parameters and operating frequency.

Across this set of designs, the FEM-recomputed mean group velocity increases with the target bandwidth. This trend indicates that maintaining both a small dispersion slope and limited group-velocity variation over a wider frequency range restricts further reduction of the mean group velocity. The surrogate captures this overall bandwidth dependence while identifying low-group-velocity structures for the different targets. For the 1.0% bandwidth design, FEM reproduces the overall dispersion profile and low mean group velocity, but the recomputed $R_v$ exceeds the constraint. Because $R_v$ depends on group-velocity extrema, it is particularly sensitive to local dispersion discrepancies. Final constraint satisfaction must therefore be determined from the FEM-recomputed results.

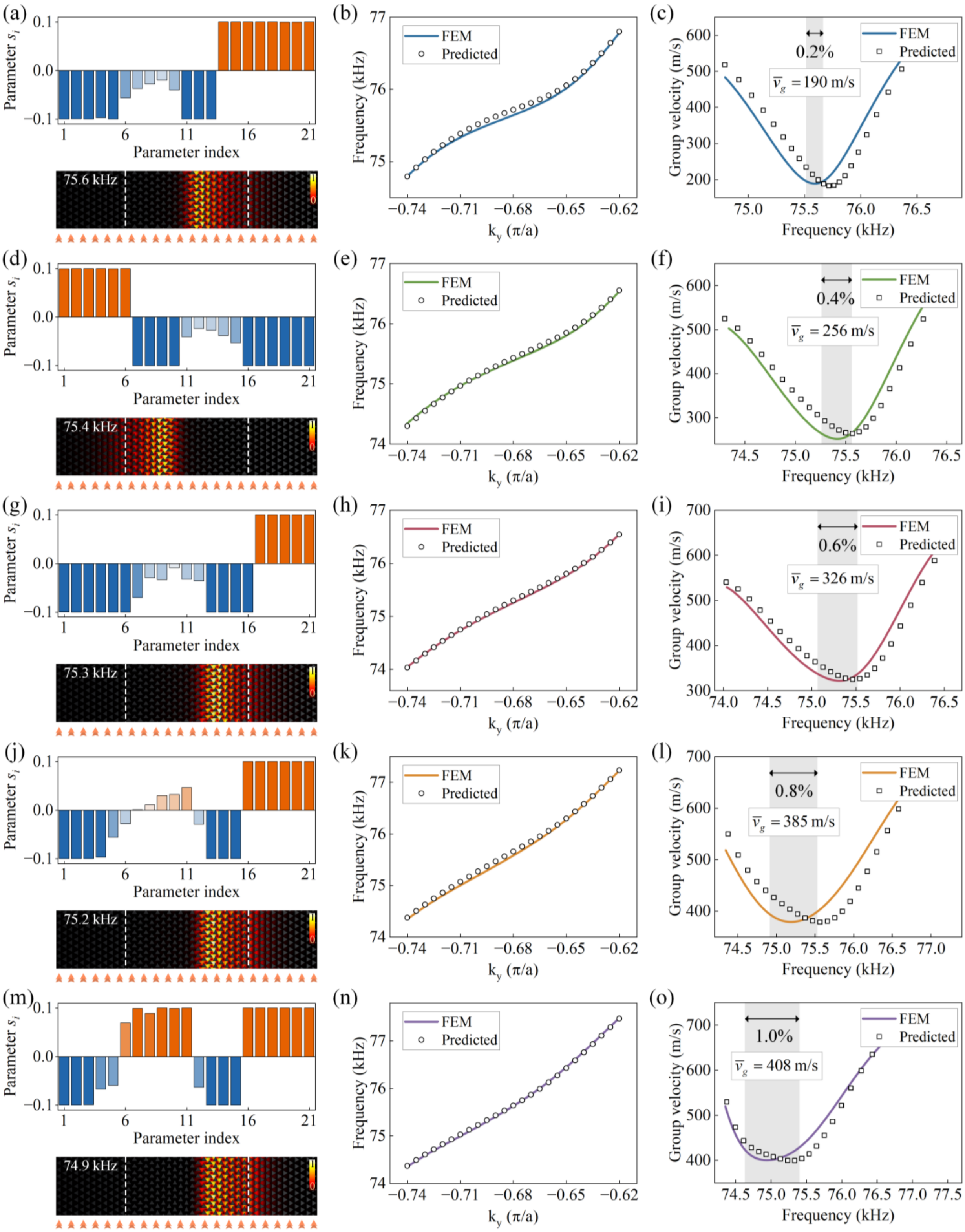

**Fig. 6.** Inverse design of low mean group velocity at different target bandwidths. From top to bottom, the rows correspond to $\beta$ = 0.2%, 0.4%, 0.6%, 0.8%, and 1.0%. Each row shows the deformation profile and plane-wave-excited normalized energy-density field, boundary-state dispersion, and group velocity. Solid lines denote FEM results, markers denote surrogate predictions, and gray regions indicate the target bandwidth windows. Field frequencies are labeled in the panels. White dashed lines mark the design-region boundaries, and orange arrows indicate the incident-wave direction.

### 5.3. Prediction errors and applicability

The two inverse-design tasks exhibit differences in predictive accuracy across dispersion profiles, with more pronounced local discrepancies in the low-mean-group-velocity designs. The surrogate reconstructs each branch from an initial frequency and cumulative positive frequency increments. Low group velocity corresponds to smaller increments between adjacent wavevector samples, so an absolute error of a given magnitude produces a larger relative deviation and affects the predicted local slope and curvature. Moreover, the training loss balances frequency, first-difference, and second-difference errors across the entire training set without separately weighting low-slope branches. Global test accuracy therefore does not imply uniform local accuracy across all dispersion profiles. Because the low-mean-group-velocity objective directly favors smaller slopes, the genetic algorithm may preferentially retain candidates whose local frequency increments are underestimated, potentially amplifying surrogate bias in low-slope regions.

Frequency errors also propagate into the group-velocity metrics through continuous dispersion reconstruction. The 25 predicted frequency samples undergo seventh-order Chebyshev fitting, evaluation on an 801-point grid, and differentiation, which can amplify existing local discrepancies. Mean group velocity represents the overall slope within the target window and is comparatively less sensitive to local errors. In contrast, $R_v$ depends on the maximum, minimum, and mean group velocities and is more sensitive to local curvature and the locations of extrema. The discrepancies observed during verification of the low-mean-group-velocity designs are consistent with the combined effects of local dispersion-prediction errors and derivative sensitivity. The surrogate is suitable for rapid evaluation and screening of candidate structures, whereas final dispersion relations and group-velocity metrics require quantitative verification through direct FEM calculations. Additional training samples near low-slope candidates or local iterative correction between the surrogate and FEM could improve predictive accuracy for these objectives.

The potential transferability of this approach lies primarily in explicitly representing the intrinsic constraints of the target response, rather than in the particular phononic crystal geometry. For other continuous spectral curves, dispersion relations, or derivative-sensitive responses, the output parameterization and loss function can be adapted to the relevant ordering, symmetry, and continuity requirements, using corresponding high-fidelity training data. The approach may therefore be extended to dispersion inverse design in other phononic crystals, acoustic and elastic metamaterials, and guided-wave structures. Such extensions require retraining and verification for the new geometries, materials, and parameter ranges. The effects of manufacturing tolerances, material losses, structural disorder, and finite-size propagation also require further investigation.

## 6. Conclusion

This study presents a supercell-enabled, constraint-preserving surrogate-assisted framework for the inverse design of boundary-state dispersion in phononic crystals with pseudomagnetic fields. Area-preserving anisotropic deformation continuously shifts the local Dirac points, while independent deformation parameters for 21 rows of unit cells within the supercell provide distributed control over the target dispersion branch. The residual surrogate preserves the ordering of predicted frequency samples and incorporates first- and second-difference losses to capture dispersion slope and curvature. Combined with genetic search and independent FEM calculations, the framework supports rapid candidate screening followed by quantitative numerical verification. By treating boundary-state dispersion itself as the direct design target, the framework extends inverse design beyond isolated frequencies or bandwidth optimization toward group-velocity control through application-oriented dispersion engineering.

The five-model ensemble achieves mean absolute errors of 4.59 Hz for frequency and 0.301 Hz for first differences on the independent test set. The framework is demonstrated through two representative prescribed-bandwidth design objectives: highly linear dispersion, quantified by group-velocity uniformity, and slow-wave dispersion, realized by minimizing the mean group velocity subject to a uniformity constraint. For the first objective, FEM verification yields relative group-velocity variations of 0.142%–1.183% over target relative bandwidths of 4%–6%. For the second, mean group velocities of 190–385 m/s are obtained over bandwidths of 0.2%–0.8%, with all four designs satisfying $R_v \leq 5\%$. At a target bandwidth of 1.0%, the FEM-recomputed mean group velocity is 408 m/s, but $R_v = 6.102\%$, exceeding the prescribed limit.

These results illustrate the trade-offs among operating bandwidth, mean group velocity, and group-velocity uniformity within the investigated design domain. Low-slope dispersion presents greater prediction challenges, and derivative-based group-velocity metrics are particularly sensitive to local errors. The surrogate is therefore suited to efficient candidate search, whereas final performance requires quantitative FEM verification. The response-constraint parameterization and surrogate-search–FEM-verification workflow may also be adapted to other continuous spectral curves and derivative-sensitive responses. Future work should incorporate experimental dispersion measurements, manufacturing tolerances, material losses, and robust inverse optimization to assess applicability to practical wave-control structures.